\documentclass[epj]{webofc}
\usepackage[utf8]{inputenc}
\usepackage[varg]{txfonts}   
\usepackage{booktabs}
\usepackage{xcolor}

\usepackage{ mathrsfs }
\usepackage{amssymb}
\usepackage{slashed}
\usepackage{color}
\usepackage{mathtools}
\usepackage{bbm}
\usepackage{pgfplots}
\usepackage{wrapfig}
\usepackage{pgfgantt}
\usepackage{caption}
\usepackage{subcaption}

\definecolor{darkred}{rgb}{0.4,0.0,0.0}
\definecolor{darkgreen}{rgb}{0.0,0.4,0.0}
\definecolor{darkblue}{rgb}{0.0,0.0,0.4}
\usepackage[bookmarks,linktocpage,colorlinks,
    linkcolor = darkred,
    urlcolor  = darkblue,
    citecolor = darkgreen]{hyperref}
\wocname{EPJ Web of Conferences}
\woctitle{Lattice2017}
\begin{document}
%
\selectlanguage{english}
\title{%
The $B_{(s)} \to D_{(s)}l\nu$ Decay with Highly Improved Staggered Quarks and NRQCD
}
\author{%
\firstname{Euan} \lastname{McLean} \inst{1}\fnsep\thanks{Speaker, \email{e.mclean.1@research.gla.ac.uk}} \and
\firstname{Christine} \lastname{Davies}\inst{1} \and
\firstname{Brian}  \lastname{Colquhoun}\inst{2} \and
\firstname{Andrew}  \lastname{Lytle}\inst{1} \\
\firstname{}\lastname{HPQCD Collaboration}
}
\institute{%
SUPA, School of Physics \& Astronomy, University of Glasgow, Glasgow, G12 8QQ, UK
\and
KEK Theory Center, High Energy Accelerator Research Organization (KEK)
Tsukuba 305-0801, Japan
}
\abstract{%
We report on progress of a lattice QCD calculation of the $B\to Dl\nu$ and $B_s\to D_s l\nu$ semileptonic form factors. We use a relativistic staggered action (HISQ) for light and charm quarks, and an improved non-relativistic (NRQCD) action for bottom, on the second generation MILC ensembles.
}
\maketitle
\section{Introduction}\label{intro}

Precise theoretical determinations of weak $B$ decay form factors are important inputs to the search for new physics in the flavour sector. For example, by combining theoretical predictions and experimental branching fractions for decays like $B_{(s)}\to D_{(s)}l\nu$, one can deduce the CKM element $|V_{cb}|$. Obtaining this to high precision is required to check the unitarity of the CKM matrix, an important test of the standard model (SM). $|V_{cb}|$ is also the dominant source of uncertainty in many current SM calculations.

There is also interest in $b \to c$ decays due to a number of anomalies (tension between experiment and the SM) in semileptonic decays involving a $b\to c$ transition. For example there are persistent discrepancies between experimentally observed values and SM predictions for the ratios $R(D^{(*)}) = \mathcal{B}(B \rightarrow D^{(*)}\tau \nu)/\mathcal{B}(B\to D^{(*)} l \nu)$ ($l=e$ or $\mu$) \cite{PhysRevLett.109.101802}.

One of the main challenges of simulating $B$ decays comes from the large $b$ quark mass.
Heavy quarks cause lattice results to develop discretization effects that grow as $(am_h)^n$, where $a$ is the lattice spacing, $m_h$ is the mass of the heavy quark, and $n$ is a positive integer. By using an improved non-relativistic effective theory (NRQCD), we can simulate the $b$ quark at its physical mass.  This framework is being used by the HPQCD collaboration in other calculations with a $b\to c$ transition, including $B\to D^*$ \cite{Harrison:2016gup}, $B_c\to \eta_c$, and $B_c\to J/\psi$ \cite{Colquhoun:2016osw}. Here, and below, we are taking $l\nu$ in the final state to be implicit. These calculations are all performed on the second generation MILC gluon configurations, building on earlier work on previous configuration sets.


\section{Methodology}
\label{sec:deets}

The form factors for a semileptonic pseudoscalar to pseudoscalar decay, $f_{0,+}(q^2)$, can be deduced from calculating the expectation value of a $b\to c$ vector operator between the appropriate meson states. The form factors and expectation values are related via
\begin{align}
\label{eq:formfactors}
\langle D_{(s)} (p) | V^{\mu} | B_{(s)} (p') \rangle  &= f^{B_{(s)}\to D_{(s)}}_{+} (q^2) \left[ p'^{\mu} + p^{\mu} - {M_{B_{(s)}}^2 - M_{D_{(s)}}^2 \over q^2} q^{\mu} \right] + f^{B_{(s)}\to D_{(s)}}_0(q^2) {M_{B_{(s)}}^2 - M_{D_{(s)}}^2 \over q^2 } q^{\mu},
\end{align}
where $|H(p)\rangle$ is a pseudoscalar meson $H$ state with momentum $p$, and $q = p' - p$. We work in a frame where the $B_{(s)}$ is at rest, so results at different $q^2$ can be achieved by varying $\vec{p}$. We introduce $\vec{p}$ using a momentum twist on the gluon fields. We fix the direction of $\vec{p}$ to be $\vec{p}=|\vec{p}|(1,1,1)$, and vary $|\vec{p}|$ accordingly.

Our calculation is performed on second generation $n_f = 2 + 1 + 1$ MILC ensembles \cite{Bazavov:2012xda}. Details of these ensembles can be found in Table \ref{table:ensembles}.
\begin{table}
\begin{center}
 \begin{tabular}{ c c c c c c c c c }
 \hline
 Set & $a[\text{fm}]$ & $am_l$ & $am_s$ & $am_c$ & $u_0$ & $L_x/a$ & $L_t/a$ & $N_{\text{cfg}}$ \\ [0.5ex] 
 \hline
 1 & 0.1474(15) & 0.013 & 0.065 & 0.838 & 0.8195 & 16 & 48 & 1020 \\ [1ex]
 2 & 0.1219(9) & 0.0102 & 0.0509 & 0.635 & 0.8340 & 24 & 64 & 1053 \\ [1ex]
 3 & 0.0884(6) & 0.0074 & 0.037 & 0.440 & 0.8525 &  32 & 96 & 1008 \\ [1ex]
 \hline
\end{tabular}
\caption{Parameters for gluon ensembles \cite{Bazavov:2012xda}.
$a$ is the lattice spacing, deduced from a study of the
$\Upsilon$-$\Upsilon'$ splitting \cite{Dowdall:2011wh}. $u_0$ is the tadpole improvement parameter as used in \cite{Dowdall:2011wh}. $L_x$ is the spatial extent and $L_t$ the
temporal extent of the lattice. Light, strange and
charm quarks are included in the sea, their masses are given in columns 3-5. $N_{\text{cfg}}$ is the number of configurations in the ensemble. We calculate propagators from 16 time sources on each configuration to increase statistics. \label{table:ensembles}}
\end{center}
\end{table}
Our calculation requires propagators for $l$(up/down),$s$(strange),$c$(charm) and $b$(bottom) valence quarks. For $l,s$ and $c$, we use the HISQ action, in which lattice artifacts are systematically removed through $\mathcal{O}(a^2)$, for masses up to and including the physical charm mass \cite{Follana:2006rc}.

It is currently very costly to perform calculations using the HISQ action for the $b$ quark, since both very fine lattices and extrapolation in $m_b$ is needed (for example in \cite{Colquhoun:2016osw}). In this study, we instead use the improved NRQCD action to generate $b$ propagators.

NRQCD propagators $G_b(\vec{x},t)$ can be computed using a recursion relation
\begin{align}
    G_b(\vec{x},t+1) = e^{-aH} G_b(\vec{x},t), 
\end{align}
with initial condition $G_b(\vec{x},0) = \phi(\vec{x})$, where $\phi(\vec{x})$ is a field of randomly generated vectors in color space of unit length. We use the $\mathcal{O}(\alpha_s v^4)$ improved NRQCD Hamiltionian:
\begin{align}
    aH_0 =& - {\Delta^{(2)}\over 2am_b} , \\
    \nonumber
    a\delta H =& - c_1 {\left(\Delta^{(2)}\right)^2 \over 8 (am_b)^3} + c_2{ i\over 8(am_b)^2} \left( \underline{\nabla}\cdot \underline{\tilde{E}} - \underline{\tilde{E}}\cdot \nabla \right) \\
    \nonumber
    & - c_3 {1\over 8(am_b)^2} \sigma\cdot \left( \underline{\nabla}\times\underline{\tilde{E}} - \underline{\tilde{E}}\times\underline{\nabla}\right) \\
    \nonumber
    & - c_4 {1\over 2am_b} \sigma\cdot \underline{\tilde{B}} + c_5 {\Delta^{(4)}\over 24 am_b} \\
    & - c_6 {\left(\Delta^{(2)}\right)^2\over 16n (am_b)^2}.
\end{align}
$\Delta^{(2,4)}$ are the second and fourth lattice covariant derivative (discretizations of $\sum_i D_i^{2,4}$). $\underline{\tilde{E}}$ and $\underline{\tilde{B}}$ are the chromoelectric and chromomagnetic fields, the expression for these in terms of gauge links are given in \cite{Gray:2005ur}. The coefficients $\{c_i\}$ are fixed by matching lattice NRQCD to continuum QCD up to 1-loop \cite{Dowdall:2011wh},\cite{Colquhoun:2014ica}.

\begin{table}
\begin{center}
 \begin{tabular}{ c c c c c c c c c c c }
 \hline
 Set & $am_l$ & $am_s$ & $am_c$ & $am_b$ & $\epsilon_{\text{Naik}}$ & $c_{1,6}$ & $c_2$ & $c_4$ & $c_5$ & $\{T\}$ \\ [0.5ex] 
 \hline
 1 & 0.013 & 0.0705 & 0.826 & 3.297 & -0.3449 & 1.36 & 1.0 & 1.22 & 1.21 & 8, 11, 14 \\ [1ex]
 2 & 0.01044 & 0.0541 & 0.645 & 2.66 & -0.2348 & 1.31 & 1.0 & 1.20 & 1.16 & 9, 12, 15 \\ [1ex]
 3 & 0.0074 & 0.0376 & 0.434 & 1.91 & -0.1172 & 1.13 & 1.29 & 1.16 & 1.11 & 15, 18, 21 \\ [1ex]
 \hline
\end{tabular}
\caption{Parameters used in our calculation. $am_l$, $am_s$ and $am_c$ are the bare masses of the light, strange and charm valence quarks, tuned in \cite{Dowdall:2011wh}, and $am_b$ is the bare mass of the valence bottom quark, tuned in \cite{Dowdall:2011wh}. The light quarks are not at physical masses ($am_l/am_s=0.2$). We expect the form factors to be relatively insensitive to this.
$\epsilon_{\text{Naik}}$ is the Naik parameter in the HISQ action \cite{Follana:2006rc}.
$\{c_i\}$ are the coefficients for the kinetic, chromoelectric and chromomagnetic terms in the NRQCD action \cite{Dowdall:2011wh},\cite{Colquhoun:2014ica}. $\{T\}$
is the set of temporal separations between source ($B_{(s)}$ creation operator) and sink ($D_{(s)}$ annihilation operator). \label{table:quarkmasses}
}
\end{center}
\end{table}

The continuum current $V_{\mu}$ can be approximated in terms of currents between HISQ and NRQCD quarks, $\{ J_{\mu}^{(i)}\}$ according to:
\begin{align}
    V_{\mu}(x) &= (1+z_{\mu}^0 \alpha_s)\left[ J_{\mu}^{(0)}(x) + J_{\mu}^{(1)}(x)\right]     \label{eq:currentcorrections} \\
    \nonumber
    & J_{\mu}^{(0)}(x) = \bar{c}(x)\gamma_{\mu} b(x) \\
    \nonumber
    & J_{\mu}^{(1)}(x) = -{1\over 2am_b} \bar{c}(x) \gamma_{\mu}\vec{\gamma}\cdot\vec{\nabla} b(x)
\end{align}
The coefficients $z_{\mu}^0$ are set by a matching procedure between the lattice NRQCD-HISQ currents and continuum QCD in \cite{Monahan:2012dq}. $J^{(0)}_{\mu}$ and $J^{(1)}_{\mu}$ are the two leading terms in a series in $\alpha_s$ and $\Lambda_{\mathrm{QCD}}/m_b$. We also calculate all other NRQCD-HISQ currents at $\mathcal{O}(\alpha_s,\Lambda_{\mathrm{QCD}}/m_b)$ to assess their size relative to the leading currents.

Additional evidence that \eqref{eq:currentcorrections} is a good approximation can be found in our $B_c\to\eta_c$ calculation \cite{Colquhoun:2016osw}. In this, form factors are deduced from vector currents from both NRQCD-HISQ currents like \eqref{eq:currentcorrections} at physical $m_b$, and HISQ-HISQ currents extrapolated to physical $m_b$. The normalization of HISQ-HISQ currents is well understood so they provide a reliable approximation to the continuum current. The study shows good agreement between the two methods (see fig. 3 of \cite{Colquhoun:2016osw}), suggesting that eq. \eqref{eq:currentcorrections} is also a good  approximation.

We obtain the expectation values of these NRQCD-HISQ currents by performing Bayesian multiexponential fits to 2- and 3-point correlation functions from our simulation - as explained in e.g. \cite{Colquhoun:2015mfa}. We use a combination of local and exponentially smeared $B_{(s)}$ and $D_{(s)}$ interpolating operators to increase statistics and increase overlap onto ground states.

We calculate expectation values of continuum vector currents at a number of different $q^2$ values. Most of the data is close to the $q^2_{\text{max}}$ side of the range, since signal/noise problems associated with a large lattice momentum is minimized there. These can be converted to form factors $f_{0,+}(q^2)$ using \eqref{eq:formfactors}. We have made a preliminary attempt at extrapolating these results to all $q^2$ by fitting the data to a BCL parameterization \cite{PhysRevD.79.013008}:
\begin{align}
    f_{0,+}(q^2) = {1\over P_{0,+}(q^2)} \sum_n a^{0,+}_n z(q^2)^n \quad,\quad z(q^2) = {\sqrt{t_+ - q^2} - \sqrt{t_+ - t_0} \over \sqrt{t_+ - q^2} + \sqrt{t_+ - t_0} },
    \label{eq:zexpansion}
\end{align}
where we take $t_0 = t_+( 1 - \sqrt{1 - t_-/t_+})$, and $t_{\pm} = (M_{B_{(s)}} \pm M_{D_{(s)}})^2$, as in \cite{Hill:2006ub}. Fitting to the data determines the free parameters $\{a^{0,+}\}$.  We truncate this at $z^2$; adding further terms have no effect on the fit. The factors $P(q^2)$ are defined by $P_{0,+}(q^2) = \left( 1 - q^2 / M_{0,+}^2\right)$, where we set $M_0=M_+=M_{B_c}\simeq 6.3~\mathrm{GeV}$. The results are insensitive to the exact value.

To account for the discretization effects, extra free parameters are added to the fit via $a_n^{0,+}\to a_n^{0,+}(1+b_n^{0,+} am_c^2)$. This is only a preliminary step towards a full treatment of the kinematic extrapolation and discretization effects that we will implement in the future.

\section{Preliminary Results}
\label{sec:results}

\begin{figure}
\centering
\begin{subfigure}{.5\textwidth}
  \centering
  \includegraphics[width=1.0\linewidth]{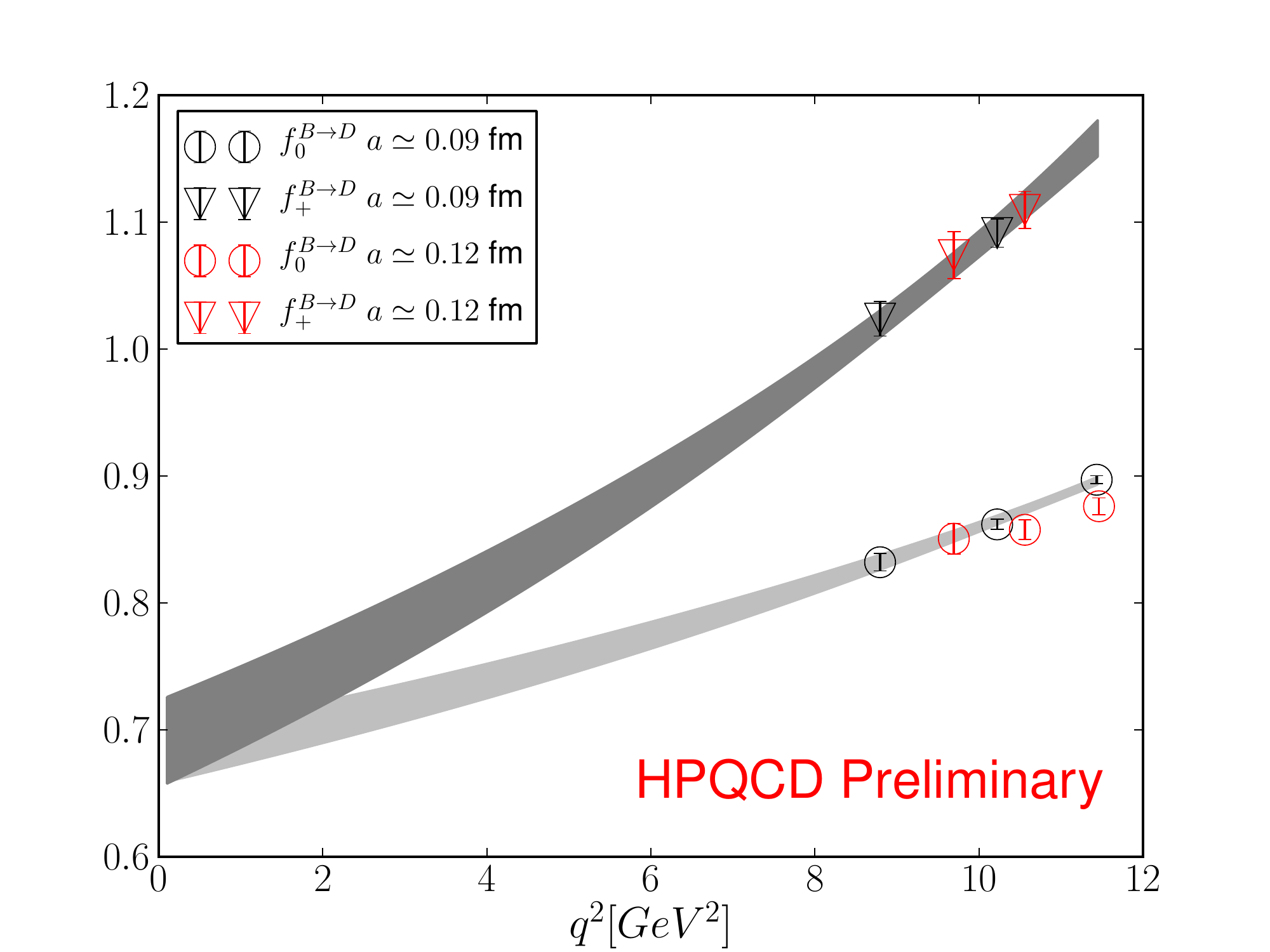}
  \label{fig:BD_ff}
\end{subfigure}%
\begin{subfigure}{.5\textwidth}
  \centering
  \includegraphics[width=1.0\linewidth]{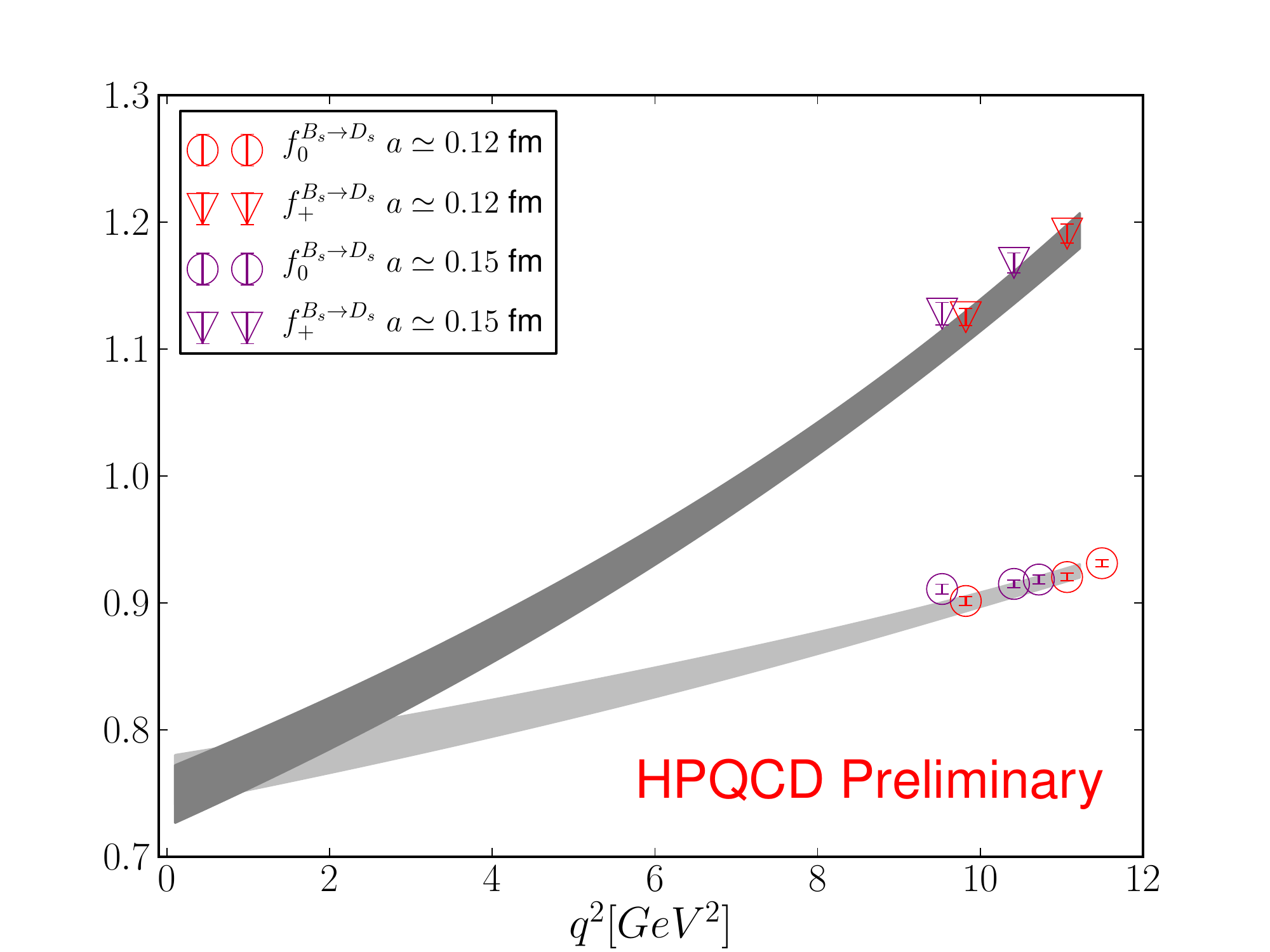}
  \label{fig:BsDs_ff}
\end{subfigure}
\caption{Form factors for the $B\to D$ (left) and $B_s \to D_s$ (right) cases. Errors on data points are statistical. The bands are produced from a fit to the data using \eqref{eq:zexpansion} as the fit form. \label{fig:ff}}
\end{figure}

Figure \ref{fig:ff} shows form factors for $B\to D$ and $B_s \to D_s$ respectively, extraced from the continuunm vector current constructed as in eq. \eqref{eq:currentcorrections}, along with our preliminary kinematic extrapolation.

The subleading lattice currents (those of $\mathcal{O}(\alpha_s,\Lambda_{\mathrm{QCD}}/m_b)$ that we neglect in our continuum current) in the temporal direction are all smaller than $\sim 5\%$ of the leading current $J^{(0)}_{0,{(s)}}$. Some subleading spatial currents however are relatively large. Namely, these are $J^{(2)}_{\mu,(s)} = - \bar{c}\gamma_\mu \gamma \cdot\overleftarrow{\nabla} b/2am_b$ and $J^{(4)}_{\mu,(s)} = - \bar{c} \overleftarrow{\nabla}_{\mu} b/2am_b$. This may cause large matching errors due to their neglection in our continuum current.

\begin{figure}
\centering
\begin{subfigure}{.5\textwidth}
  \centering
  \includegraphics[width=1.0\linewidth]{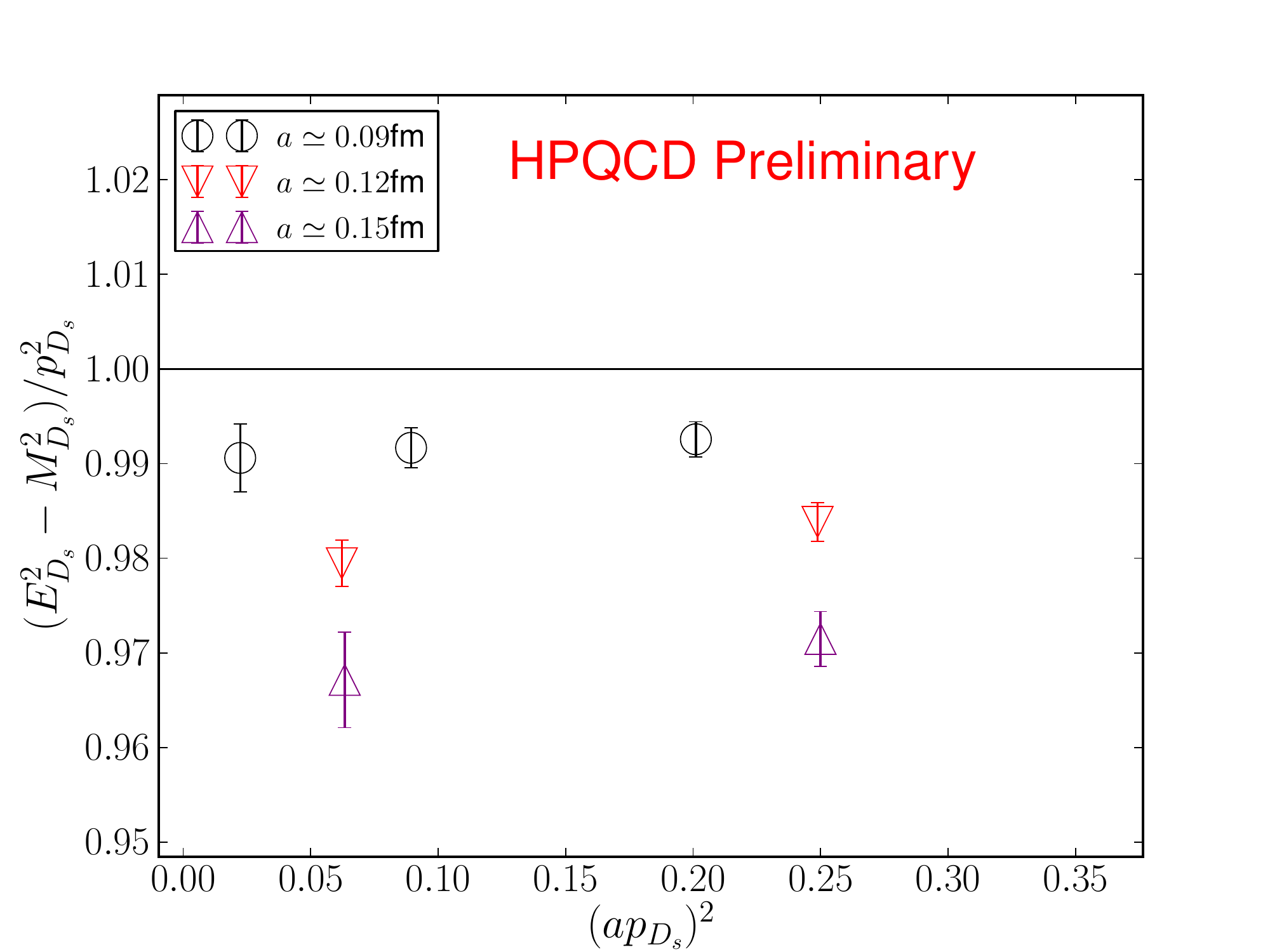}
  \label{fig:sub1}
\end{subfigure}%
\begin{subfigure}{.5\textwidth}
  \centering
  \includegraphics[width=1.0\linewidth]{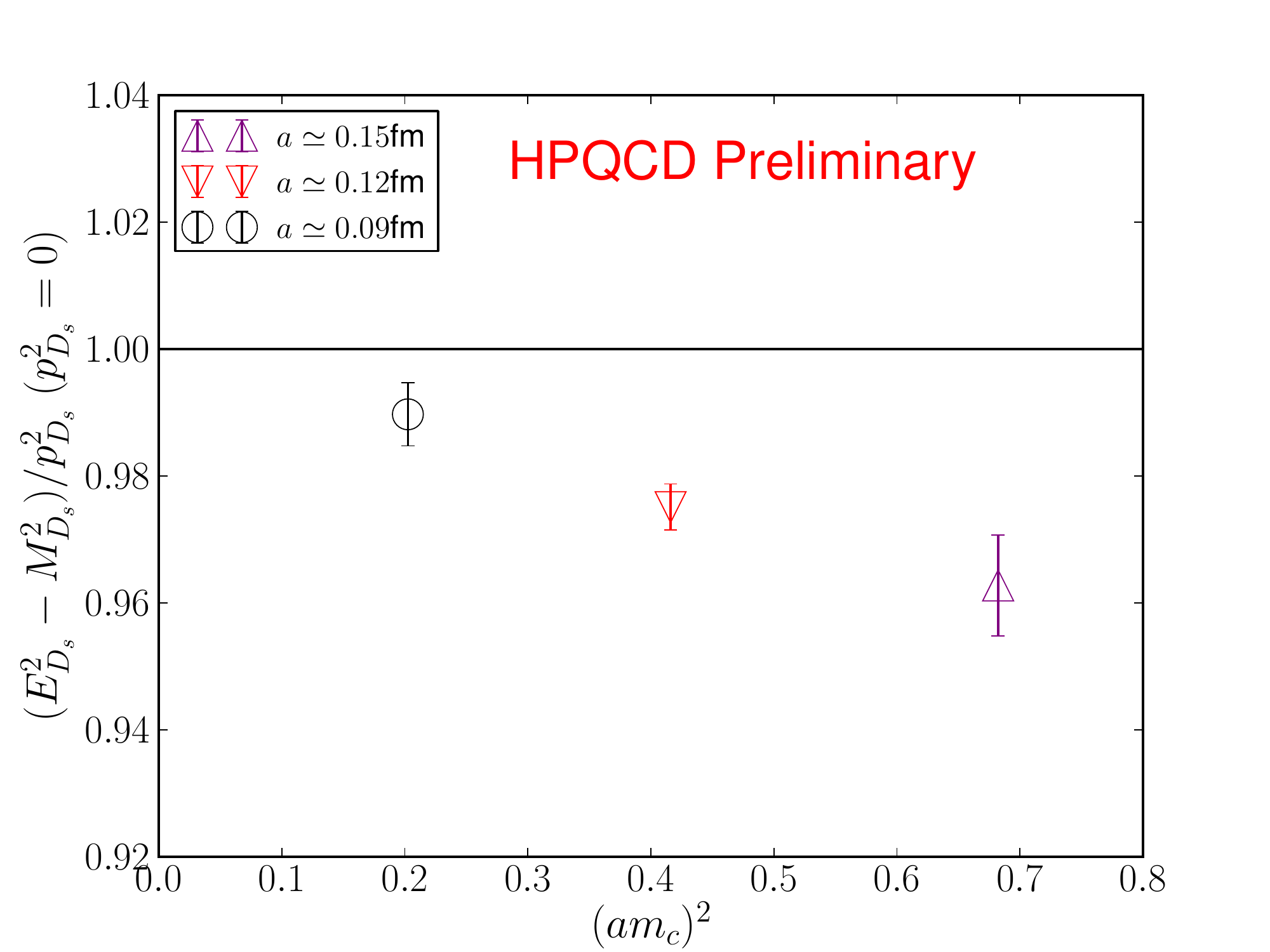}
  \label{fig:sub2}
\end{subfigure}
\caption{Left: The speed of light against $D_s$ momentum. Discretization errors cause this to differ from unity. Right: On each ensemble, the speed of light at $\left\vert a\vec{p}_{D_s} \right\vert=0$ is extrapolated from the $a\vec{p}_{D_s}\neq 0$ data (below $\left(a\vec{p}_{D_s}\right)^2 =0.5$), and plotted here. As can be seen, in the limit $am_c\to 0$, the speed of light tends towards unity, showing that discretization effects are controllable below $\left(a\vec{p}_{D_s}\right)^2 = 0.5$.}
\label{fig:speedoflight}
\end{figure}

A useful check of how well the discretization errors can be controlled as we add larger $|a\vec{p}_{D_{(s)}}|$ is to compute the speed of light $c = (E_{D_s}(\vec{p}_{D_s})^2 - M_{D_s}^2) / \vec{p}_{D_s}^2$ using our data. We can show that $c^2$ tends to unity in the continuum limit ($am_c,\left\vert a\vec{p}_{D_s} \right\vert\to 0$), see fig. \ref{fig:speedoflight} (see also \cite{Donald:2012ga}). 

\section{Conclusion}
\label{sec:future}

We are calculating form factors for the $B\to D$ and $B_s \to D_s$ semileptonic decays. This work will add to the preexisting body of work on $b\to c$ transitions by the HPQCD collaboration and will demonstrate the power and limits of our current methods. Along with current and upcoming experimental results for $B_{(s)}\to D_{(s)} l\nu$ branching fractions, our calculation will help towards better determinations of $|V_{cb}|$, and may help towards further understanding the current tensions between theory and experiment in the $B\to D$ channel.

We are currently investigating the possibility of reliably using data at large $\left\vert a\vec{p}_{D_{(s)}} \right\vert$. This will require a more sophisticated approach to dealing with the expected growth of discretization effects and a better understanding of the matching errors. In addition to the systematic issues, the signal/noise ratio shrinks exponentially as $\vec{p}_{D_{(s)}}$ is decreased. New innovations will be required to deal with this.

We are also investigating ways to exploit the Ward identity between vector and scalar currents to fix the renormalization of our continuum vector current.

Finally, we plan on computing the above form factors using a complementary HISQ $b$ approach as in \cite{Colquhoun:2016osw}, in order to test the systematic errors associated with NRQCD.

\section{Acknowledgements}

This work was performed on the Darwin supercomputer at the University of Cambridge, part of STFC’s DiRAC facility.

\clearpage
\bibliography{lattice2017}

\end{document}